\documentclass[useAMS,usenatbib]{mn2e}
\usepackage{amsmath}
\usepackage{times}
\usepackage{graphicx}

\title[X-ray Decay in SN 1978K]
{The X-ray Decay of the ultraluminous supernova SN 1978K in NGC 1313}

\author[Zhao et al.]{Hai-Hui Zhao$^{1}$, Shan-Shan Weng$^{1}$\thanks{E-mail: wengss@njnu.edu.cn}, C.-Y. Ng$^{2}$\\
$^1$\,Department of Physics and Institute of Theoretical Physics,
Nanjing Normal University, Nanjing 210023, China \\
$^2$\, Department of Physics, The University of Hong Kong, Pokfulam Road, Hong Kong \\
}

\date{}

\pagerange{\pageref{firstpage}--\pageref{lastpage}} \pubyear{2016}

\begin{document} \maketitle

\begin{abstract}

The Type IIn supernova (SN) 1978K in the nearby galaxy NGC 1313 has
remained bright in multiwavelengths for more than $\sim 25$ yr. The
archival data of SN 1978K collected with {\it ROSAT}, {\it ASCA}, {\it
XMM--Newton} and {\it Chandra} from 1990 to 2006 show no significant variation
of the soft X-ray emission but a hint of flux decrease in hard X-rays. In this
work, we perform a detailed analysis using more than 15 yr of {\it
XMM--Newton} observations. Both the 0.5--2 keV and 2--10 keV light-curves decline
as $t^{-1}$ from 2000 to 2015. The transition of light curve profiles can be
explained in a way that the reverse shock was radiative at an early phase and
then became adiabatic at late times. Such a scenario is also supported by
the spectral analysis results. We also found a decrease in the absorption
column density, which indicates the presence of a cool shell during the
radiative phase.

\end{abstract}


\section{Introduction}

Progenitors of supernovae (SNe), massive stars lose mass via stellar winds,
forming a circumstellar medium (CSM). After the explosion, the SN shock waves
expand into the surrounding CSM \citep[see reviews by ][]{smith14}. The
subsequent interaction creates a fast-forward and a reverse shock \citep[e.g.
][]{chevalier94, chatzopoulos12}. Such interaction can be investigated with
radio, infrared, ultraviolet, and X-ray observational data \citep[e.g.
][]{chugai07, tanaka12, milisavljevic13, ng13, gezari15, ryder16}. If the SNe
propagate in the steady stellar winds, we would expect that the X-ray
luminosity resulting from shocks declines in a power-law way. Recently,
\cite{dwarkadas12} collected the X-ray data of 42 SNe and found diverse 
light-curve profiles. There is increasing evidence that the density structure of
the ambient medium may not follow $r^{-2}$ as expected according to the steady
wind theory \citep{chevalier03,chandra09}. It can even be inhomogeneous or
clumpy.

SN 1978K was first discovered as a strong H$\alpha$ source in 1990 during a
spectrophotometric survey of the H~\textsc{ii} region in nearby galaxies, and it
was originally classified as a nova \citep{dopita90}. Investigating the
multiwavelength archival data, \cite{ryder93} suggested that it was actually a
Type II SN that exploded around 1978 May 22 \citep{montes97}. Based on
the X-ray and radio data, \cite{schlegel99} suggested that the behaviour of SN
1978K is typical of the `IIn' subclass of Type II SNe, which was first
designated by \cite{schlegel90}. The powerful non-thermal radio emission was
first revealed with the Molonglo Observatory Synthesis Telescope data taken in
1982 \citep{ryder93}. It reached a peak luminosity around the late 1980s, and
then declined monotonically up to now \citep{smith07}. On 1980 January 2,
the {\it Einstein} X-ray satellite observed the region of SN 1978K, and
obtained an upper limit of unabsorbed flux in 0.2--2.4 keV of $1.1 \times
10^{-13}$ erg cm$^{-2}$ s$^{-1}$. X-ray emission of SN 1978K was uncovered with
the {\it ROSAT} Position Sensitive Proportional Counter in 1990 \citep{ryder93,
lenz07}. The derived unabsorbed flux was of $\sim 10^{-12}$ erg cm$^{-2}$
s$^{-1}$ in the 0.2--2.4 keV range, that is $\sim 3 \times 10^{39}$ erg
s$^{-1}$ at a distance of 4.61 Mpc \citep{gao15}, identifying SN 1978K as an
ultraluminous X-ray source (ULX). Follow-up X-ray observations made with {\it
ASCA}, {\it XMM--Newton} and {\it Chandra} indicated that the X-ray luminosity
of SN 1978K remains bright even at $\sim 37$ yr after the explosion
\citep[e.g. ][]{lenz07, smith07}.

Since SN 1978K is luminous in multiwavelengths, the long-term broad-band
monitoring data provide a good opportunity for the study of the interaction
between the outgoing shock and the CSM. The X-ray data of SN 1978K prior to
2006 have been analysed; however, whether its flux has remained constant since the
first detection by {\it ROSAT} or started to decline was still under debate
\citep[e.g. ][]{lenz07, smith07}. It is worth noting that there are another two
ULXs (X-1 and X-2) in NGC 1313. Thus, this region has been frequently visited
by X-ray missions in order to explore the spectral evolution of these sources
\citep[e.g. ][]{feng06,bachetti13,weng14}. In the past decade, some additional
long observations were carried out with {\it XMM--Newton} (Table \ref{log}),
which is a focusing X-ray telescope having a large effective area in the soft
X-ray band. In this paper, we study the {\it XMM--Newton} observations covering
the period from 2000 to 2015 (spanning 5277 d), aiming to explore the evolution in X-ray
emission of SN 1978K at very late times. The {\it XMM--Newton} data reduction is
described in the next section, and the results are presented in Sections 3.
Discussion and conclusion follow in Section 4.

\section{Data Reduction}

The detailed study presented here was on data from the main camera onboard {\it
XMM--Newton}, the European Photon Imaging Camera (EPIC), which consists of two
MOS arrays and one PN CCD array \citep{struder01,turner01}. We analyse all of the
available {\it XMM--Newton} observations of SN 1978K made from 2000 October to
2015 March; however, we discard those data that were affected by the strong
background flares or when the source falls in the CCD gaps. The observational
details of the data used in this paper are listed in Table \ref{log}.
The {\it XMM--Newton} observations of SN 1978K are quite clean except
two short observations in 2003 December (ObsIDs = 0150280401, 0150280501), and
only less than 12\% of data on average are excluded. The data are reduced with
the Science Analysis System software (\textsc{sas}) version 14.0.0. We create the light
curve above 10 keV, and exclude background flares with a count-rate cut-off
criterion. The data from good time intervals are filtered by setting
FLAG = 0 and PATTERN $\leq$ 4 for the PN data, and PATTERN $\leq$ 12 for the
MOS data to ensure the best quality spectrum \citep{snowden16}. The source
photons are extracted from a circle aperture with a radius of 30 arcsec,
while the background is taken from the same CCD chips as the source and at a
similar distance from the readout node. Since the source is relatively faint
($flux < 10^{-12}$ erg cm$^{-2}$ s$^{-1}$), the pile-up effect can be
neglected. We produce the response files with the \textsc{sas} tasks \texttt{rmfgen} and
\texttt{arfgen} to facilitate subsequent spectral analysis. The task
\texttt{specgroup} is carried out to rebin all spectra to have at least 20
counts per bin and not to oversample the instrument energy resolution by more
than a factor of 3.

We fit the spectra in the 0.5--10 keV range using \textsc{xspec} with a 2\%
systematic error added \citep{smith16}. When available, the PN and MOS spectra
for each observation are fitted together. To account for the
cross-normalization issues, we include a multiplicative constant in the model
and fix it to unity for the PN data, but allow it to vary for the MOS data. Due
to the short exposure time, it is difficult to constrain the parameters for the
individual observations carried out before 2006. Since there is no
significant spectral evolution within less than one year, we merge some
successive observations, which are referred to as Groups 1--8 (Tables \ref{log}
and \ref{vmekal_fits}), to better constrain the model parameters.

The X-ray emission from SN 1978K can be described by an absorbed
two-temperature optically thin thermal plasma model, in which the soft and hard
components are attributed to the reverse and forward shocks, respectively.
\cite{smith07} suggested that the X-ray spectra of SN 1978K can be fitted with a
double \texttt{vmekal} model \citep[see also ][]{schlegel04} with very large
helium abundance. Here, we first apply a two-temperature non-equilibrium
ionization model (\texttt{vnei} in  \textsc{xspec}), and confirm that helium abundance
in both components needs to be large. The ionization time-scale for the soft
component is larger $3\times10^{13}$ s cm$^{-3}$, indicating that the reverse
shock collides with very dense clouds and has already reached ionization
equilibrium. Meanwhile, the ionization time-scale for the hard component is
$\sim (5$-$10)\times10^{10}$ s cm$^{-3}$, indicating a non-equilibrium condition.
Therefore, we adopt the \texttt{tbabs*(vmekal+vnei)} model in the following
spectra analysis, since \texttt{vnei} reduces to \texttt{vmekal} in ionization
equilibrium and the latter is much faster to compute.

The best-fitting spectral parameters with uncertainties at 90\% confidence level
are listed in Table \ref{vmekal_fits}. Both the absorbed and unabsorbed fluxes
are calculated with the convolution model \texttt{cflux}. The evolution of
the X-ray luminosity and the equivalent hydrogen column density are not
sensitive to the spectral models, while the parameters of two shocks, including
plasma temperatures and large He abundances, generally agree with those by
\cite{smith07}, who also gave an all-round analysis of the {\it XMM-Newton}
data prior to 2006. However, there is no evidence for the large He abundances
detected in the {\it Chandra} data (Schlegel et al., in preparation). The
divergence might be due to the calibration differences between {\it Chandra}
and {\it XMM--Newton} \citep[e.g. ][]{nevalainen10,schellenberger15}. Note that
our absorption column density [$N_{\rm H} \sim (2.1-2.5) \times 10^{21}$
cm$^{-2}$] is slightly higher than the values reported in \cite{smith07} and
\cite{lenz07}, due to the choice of the updated solar abundances by
\cite{wilm00}. This value is also much larger than the absorption in the
direction of NGC 1313, $\sim 3.7 \times 10^{20}$ cm$^{-2}$ \citep{schlegel98}.


\begin{table*} \small
\caption{{\it XMM--Newton} observations of SN 1978K. Net exposure: clean exposure
after background flares excluded.} \label{log}
\centering \begin{tabular}{c c c c c} \hline
Group & ObsID & Obs date & Instruments & Net exposure (ks) \\
\hline
Group1 & 0106860101 & 2000-10-17 & PN/MOS1/MOS2 & 24.1/28.7/28.7 \\
\\
Group2 & 0150280301 & 2003-12-21 & PN/MOS1/MOS2 & 7.7/10.5/10.5  \\
  Group2   & 0150280401 & 2003-12-23 & PN/MOS1/MOS2 & 3.3/8.1/8.5    \\
  Group2   & 0150280501 & 2003-12-25 & PN/MOS1/MOS2 & 6.4/9.3/9.4    \\
   Group2 & 0150280601 & 2004-01-08 & PN/MOS1/MOS2 & 8.4/12.8/13.0  \\
  Group2  & 0150281101 & 2004-01-16 &    MOS1/MOS2 & 8.5/8.5        \\
  Group2  & 0205230201 & 2004-05-01 &    MOS1/MOS2 & 8.7/9.3        \\
\\
Group3 & 0205230301 & 2004-06-05 & PN/MOS1/MOS2 & 8.9/11.6/11.5  \\
 Group3    & 0205230401 & 2004-08-23 & PN/MOS1/MOS2 & 9.0/14.9/15.2  \\
 Group3    & 0205230501 & 2004-11-23 &    MOS1/MOS2 & 15.5/15.5      \\
 Group3    & 0205230601 & 2005-02-07 & PN/MOS1/MOS2 & 9.8/12.6/12.9  \\
\\
Group4 & 0301860101 & 2006-03-06 & PN/MOS1/MOS2 & 17.5/21.3/21.3 \\
 Group4     & 0405090101 & 2006-10-15 & PN/MOS1/MOS2 & 86.0/119.0/119.6\\
\\
Group5 & 0693850501 & 2012-12-16 & PN/MOS1/MOS2 & 97.9/117.7/120.2\\
Group5     & 0693851201 & 2012-12-22 & PN/MOS1/MOS2 & 101.4/121.1/123.7\\
\\
Group6 & 0722650101 & 2013-06-08 & PN/MOS1/MOS2 & 20.0/30.1/30.1 \\
\\
Group7 & 0742590301 & 2014-07-05 & PN/MOS1/MOS2 & 53.6/61.0/61.1 \\
\\
Group8 & 0742490101 & 2015-03-30 & PN/MOS1/MOS2 & 84.8/98.0/100.1 \\
\hline

\end{tabular}

\end{table*}

\section{Results}

\begin{figure}
\includegraphics[width=9cm]{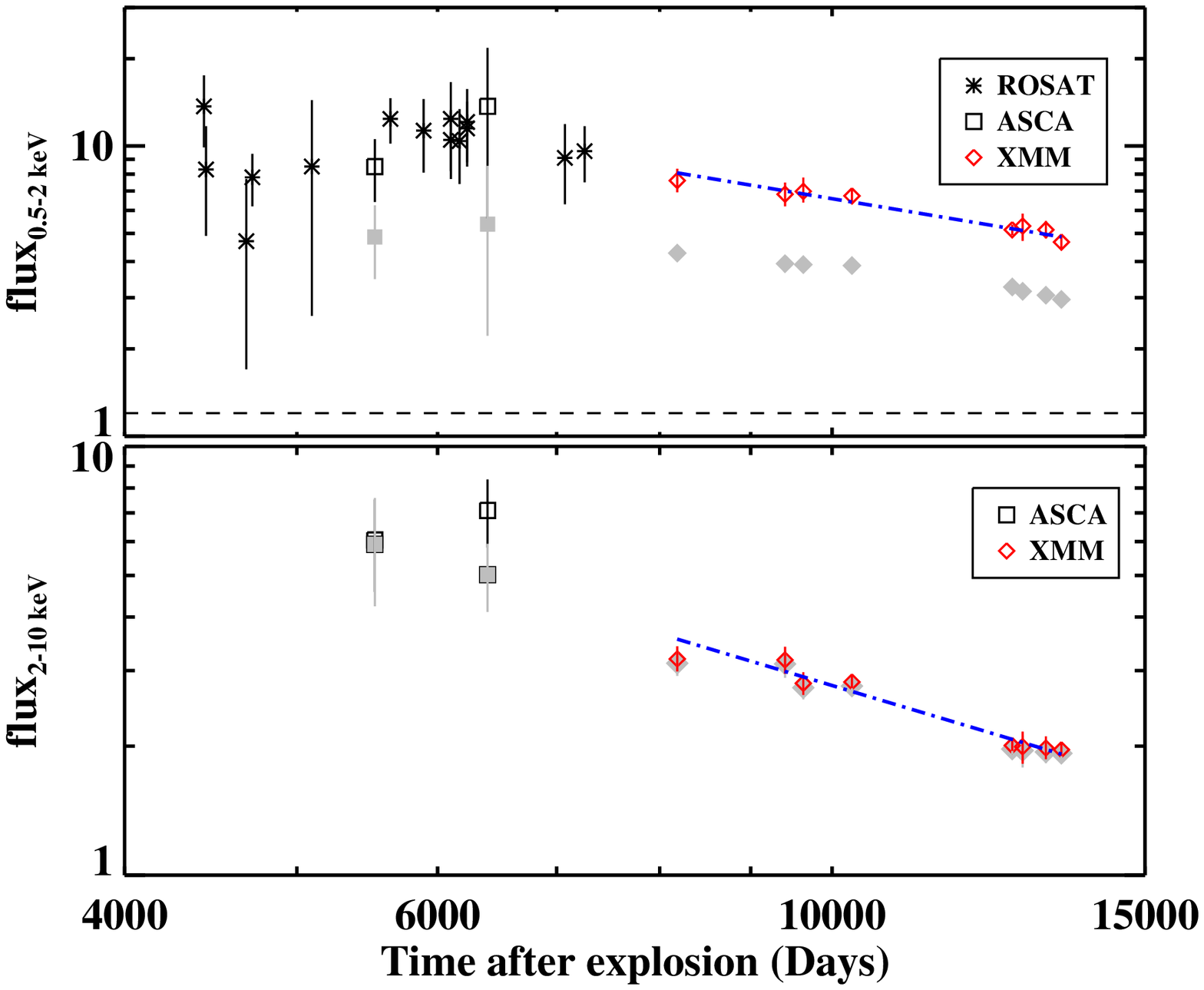}
\caption{0.5--2 keV (upper panel) and 2--10 keV (bottom panel) X-ray fluxes in
units of 10$^{-13}$ erg cm$^{-2}$ s$^{-1}$ versus SN age. The black data points
mark the unabsorbed fluxes from other X-ray missions \citep[adopted from ][and
references therein]{schlegel04}, and grey solid symbols correspond to the
absorbed fluxes. The horizontal dashed line in the upper panel represents the
upper limit derived from the {\it Einstein} IPC observation on 1980 January 2
(age $\sim$ 590 d). The red points are from this paper (Table
\ref{vmekal_fits}), which can be fitted by a power-law decay (blue dot--dashed
lines). \label{lightcurve}}
\end{figure}

Since the first {\it XMM--Newton} observation carried out on 2000 October 17,
the X-ray flux of SN 1978K had dropped by 40\% in the last 15 yr. Assuming a
distance of 4.61 Mpc \citep{gao15}, the unabsorbed luminosities in the 0.5--10 keV
band are 2.96 $\times 10^{39}$  and 1.76 $\times 10^{39}$ erg
s$^{-1}$ on 2000 October 17 and 2015 March 30. The detailed flux evolution is
plotted in Fig.\ref{lightcurve}. It is quite obvious that the unabsorbed
X-ray fluxes of SN 1978K display a power-law decay. Using the \textsc{idl} routine
\texttt{linfit}, we fit the light curves with the following function:
\begin{equation}
\log(flux) = A(0) \times \log(t-t_{0})+A(1),
\end{equation}
where $t$ is the observational date in days, $t_0$ is taken to be MJD 43650
\citep[1978 May 22; ][]{montes97} when the SN was at peak luminosity and $A(0)$
and $A(1)$ are the fitting parameters. The derived values of $A(0)$ are $-1.02
\pm 0.15$ for 0.5--2 keV, and $-1.24 \pm 0.10$ for 2--10 keV. In Fig.
\ref{lightcurve}, we also plot the X-ray fluxes measured by {\it Einstein},
{\it ROSAT} and {\it ASCA} \citep[][and references therein]{schlegel04}. These
archival data indicate that the X-ray flux in the 0.5--2 keV band had a shallower
decay or remained constant at an earlier stage. On the other hand, the 2--10 keV
flux estimated from the {\it ASCA} data could be consistent with the extrapolated
value of our power-law fitting (bottom panel of Fig. \ref{lightcurve}).

As seen in Table \ref{vmekal_fits}, the absorption column density ($N_{\rm H}$)
decreases with time. It is worth noting that the parameters in Groups 1--4 (and
5--8) vary by less than 20\%. Therefore, in order to increase the
signal-to-noise ratio, we divide all spectra into two parts, those collected
before 2006 October 16 and those collected between 2012 and 2015. The same spectral model is then applied to these two sets of spectra (Table \ref{combined_fits}),
respectively. We found that the value of $N_{\rm H}$ has decreased by more than
5$\sigma$ [in the range (2.53--2.12) $\times 10^{21}$ cm$^{-2}$] over the last $\sim$10 yr. However, we can not estimate the evolution of the parameters
of the hot plasma components because of their large uncertainties.


\begin{table*} \scriptsize
\caption{Best-fitting spectral parameters of SN 1978K with the model of
\texttt{tbabs*(vmekal+vnei)}. } \label{vmekal_fits}
\centering \begin{tabular}{l c c c c c c c c c c c} \hline ObsID & $N_{\rm H}$
& $kT_{\rm soft}$ & He$_{\rm soft}$ & $kT_{\rm hard}$ & He$_{\rm hard}$
& $\log(\tau_{\rm hard})$ &$flux_{\rm 0.5-2 keV}^{\sharp}$ & $flux_{\rm 2-10 keV}^{\sharp}$ & $\chi^2$/dof \\
  & ($10^{21}$ cm$^{-2}$) & (keV) &  & (keV) &  & (s cm$^{-3}$) & ($10^{-13}$ erg cm$^{-2}$ s$^{-1}$) & ($10^{-13}$ erg cm$^{-2}$ s$^{-1}$) & \\
\hline
Group1 & $2.70_{-0.38}^{+0.39}$ & $0.63_{-0.04}^{+0.03}$ & $29_{-9}^{+12}$  & $4.0_{-0.8}^{+1.2}$ & $49_{-22}^{+54}$ & $10.8_{-0.2}^{+0.1}$  & $4.27_{-0.09}^{+0.09}/7.6_{-0.7}^{+0.7}$ & $3.11_{-0.20}^{+0.22}/3.19_{-0.20}^{+0.22}$ & 129.5/165 \\
Group2 & $2.54_{-0.38}^{+0.38}$ & $0.60_{-0.03}^{+0.03}$ & $29_{-9}^{+10}$  & $4.7_{-1.0}^{+1.4}$ & $22_{-16}^{+37}$ & $11.1_{-0.2}^{+0.3}$  & $3.93_{-0.10}^{+0.10}/6.8_{-0.6}^{+0.7}$ & $3.11_{-0.22}^{+0.23}/3.17_{-0.22}^{+0.23}$ & 467.4/431 \\
Group3 & $2.65_{-0.35}^{+0.43}$ & $0.57_{-0.05}^{+0.03}$ & $33_{-10}^{+12}$ & $3.7_{-0.6}^{+0.8}$ & $64_{-34}^{+102}$& $10.9_{-0.3}^{+0.2}$  & $3.90_{-0.08}^{+0.08}/7.0_{-0.6}^{+0.8}$ & $2.73_{-0.16}^{+0.17}/2.80_{-0.16}^{+0.17}$ & 413.9/401 \\
Group4 & $2.52_{-0.22}^{+0.23}$ & $0.59_{-0.03}^{+0.02}$ & $30_{-6}^{+7}$   & $3.7_{-0.3}^{+0.4}$ & $106_{-40}^{+67}$ & $10.7_{-0.2}^{+0.2}$  & $3.87_{-0.04}^{+0.04}/6.7_{-0.4}^{+0.4}$ & $2.76_{-0.09}^{+0.09}/2.82_{-0.09}^{+0.09}$ & 450.6/413 \\
Group5 & $2.06_{-0.16}^{+0.15}$ & $0.63_{-0.01}^{+0.01}$ & $24_{-4}^{+4}$   & $3.4_{-0.3}^{+0.3}$ & $42_{-31}^{+41}$ & $11.1_{-0.2}^{+0.5}$ &  $3.26_{-0.03}^{+0.03}/5.1_{-0.2}^{+0.2}$ & $1.96_{-0.05}^{+0.05}/2.00_{-0.05}^{+0.05}$ & 507.6/481 \\
Group6 & $2.32_{-0.47}^{+0.39}$ & $0.63_{-0.03}^{+0.03}$ & $30_{-13}^{+12}$ & $3.6_{-0.9}^{+1.3}$ & $9_{-6}^{+246}$ & $11.5_{-0.3}^{+2.2 }$  & $3.15_{-0.08}^{+0.08}/5.3_{-0.6}^{+0.5}$ & $1.95_{-0.17}^{+0.16}/1.99_{-0.17}^{+0.16}$ & 169.1/143 \\
Group7 & $2.35_{-0.28}^{+0.28}$ & $0.65_{-0.02}^{+0.02}$ & $26_{-6}^{+7}$   & $4.1_{-0.7}^{+1.3}$ & $6_{-4}^{+19}$  & $11.6_{-0.4}^{+2.1 }$  & $3.06_{-0.05}^{+0.05}/5.1_{-0.3}^{+0.4}$ & $1.93_{-0.11}^{+0.12}/1.97_{-0.11}^{+0.12}$ & 208.1/177 \\
Group8 & $2.08_{-0.23}^{+0.23}$ & $0.61_{-0.02}^{+0.02}$ & $20_{-5}^{+5}$   & $3.4_{-0.3}^{+0.4}$ & $35_{-15}^{+28}$ & $11.1_{-0.2}^{+0.2}$  & $2.96_{-0.04}^{+0.04}/4.7_{-0.3}^{+0.3}$ & $1.92_{-0.06}^{+0.06}/1.96_{-0.06}^{+0.07}$ & 214.3/236 \\
\hline
\end{tabular}\\
{\footnotesize $^a$Both absorbed (former) and unabsorbed
(latter) fluxes are calculated in  0.5--2 keV and 2--10 keV, respectively.
All errors are in the 90\% confidence level. }
\end{table*}


\begin{table*} \scriptsize
\centering \begin{tabular}{l c c c c c c c c c c c} \hline Date & $N_{\rm H}$ &
$kT_{\rm soft}$ & He$_{\rm soft}$ & norm$_{\rm soft}$ & $kT_{\rm hard}$ &
He$_{\rm
hard}$ & $\log(\tau_{\rm hard})$ & norm$_{\rm hard}$ & $\chi^2$/dof \\
  & ($10^{21}$ cm$^{-2}$) & (keV) &  & ($\times 10^{-4}$) &(keV) & & (s cm$^{-3}$) &($\times 10^{-4}$)  &   & \\
\hline
2000-2006 &  $2.53_{-0.15}^{+0.15}$ & $0.60_{-0.01}^{+0.01}$ & $29_{-4}^{+4}$ & $1.21_{-0.06}^{+0.06}$ & $3.9_{-0.3}^{+0.3}$ & $68_{-21}^{+33}$ & $10.9_{-0.1}^{+0.1}$ & $0.19_{-0.06}^{+0.08}$  & 1643.5/1437\\
2012-2015 &  $2.12_{-0.11}^{+0.11}$ & $0.63_{-0.01}^{+0.01}$ & $23_{-3}^{+3}$ & $0.99_{-0.04}^{+0.04}$ & $3.5_{-0.2}^{+0.2}$ & $41_{-21}^{+22}$ & $11.1_{-0.1}^{+0.3}$ & $0.23_{-0.07}^{+0.21}$  & 1257.7/1064\\
\hline
\end{tabular}
\caption{Same as Table \ref{vmekal_fits} but with spectra regrouped.}
\label{combined_fits}
\end{table*}

\section{Discussion and conclusion}

Multiband observations of Type II SNe bear the imprint of interaction between
the SNe and CSM ejected by their progenitors. For the non-radiative case, we would
expect that the X-ray emission decreases inversely with time as $t^{-1}$ if SNe
expand into a steady wind medium \citep[see e.g. ][]{immler03}.
\cite{dwarkadas12} investigated the light curves of 42 SNe, but found that none of
them declines as $L_{\rm X} \propto t^{-1}$. Before our work, \cite{lenz07}
claimed that the soft X-ray flux of SN 1978K remained at a constant level,
whereas the hard X-ray emission dropped slightly. In the meantime, the radio
light curves had faded steadily since 1991 \citep{smith07}
\footnote{https://www.ast.cam.ac.uk/ioa/wikis/SEES/images/f/f1/Ryder.pdf}.
Here, we analyse all archival {\it XMM--Newton} data for both light-curve and
spectral evolution. It is remarkable that both the soft and hard X-ray light
curves turn into a $t^{-1}$ decline around the year 2000 ($\sim 6000 - 8000$ d
 after explosion). This result implies that the shock wave is now
propagating through the medium with a density gradient $r^{-2}$
\citep{chevalier03,chandra09,dwarkadas12}, and the free--free radiation
dominates the X-ray emission. Alternatively, the transition of the reverse shock
from radiative to adiabatic phases could account for the change of the light
curve decay rate. As we discuss below, such a scenario is supported by both the
temporal and spectral evolutions, and it was proposed to explain the behaviour of
SN 1993J \citep{chandra09}.

SN 1993J is another bright SN and is well studied in broad-band wavelengths
\citep[e.g. ][ and references therein]{bjornsson15,jerkstrand15}. During the
first $\sim 5$ yr, the soft and hard X-ray fluxes had shallow decays at
different rates, and then both underwent a $t^{-1}$ decline at later times
\citep{chandra09}, analogous to what SN 1978K exhibits (Fig.\ref{lightcurve}). 
With the self-similar assumption of a steady progenitor wind,
the free--free emission from the adiabatic shock is suggested to follow the time
evolution of $t^{-1}$ \citep[e.g. ][]{chevalier03}. The forward shock,
corresponding to the harder component, has been in the adiabatic phase since the very
beginning. On the other hand, the reverse shock (the softer component) could be
radiative at an early stage, having a flatter profile of soft X-ray light
curves, and then it becomes adiabatic after a cooling time of the gas behind the
reverse shock \citep{chandra09}.

The caution is that SN 1993J is categorized as a Type IIb SN \citep{woosley94},
while SN 1978K is of Type IIn. SN 1978K is still barely resolved in the new very 
long baseline interferometry image at 8.4GHz on 2015 March 29; therefore, 
\cite{ryder16} suggested that its average expansion velocity in the last 37 yr 
had been less than 1500 km s$^{-1}$. Compared to the case in SN 1993J, the much 
slower shock velocity of $\sim$500--600 km s$^{-1}$ in SN 1978K shown by the optical emission lines indicates that the SN blast wave has been decelerated by the very dense CSM \citep{chugai95,kuncarayakti16}. Since the cooling time of the gas behind the
reverse shock is anti-correlated with the shock velocity \citep[$t_{\rm c}
\propto v_{\rm s}^{-5.2}$; ][]{chevalier94,chandra09}, the lower shock speed in
SN 1978K results in a longer transition time ($\sim$ 6000--8000 d;
Fig. \ref{lightcurve}) for radiative cooling. In addition, the large He
abundances observed in SN 1978K (Table \ref{combined_fits}) would extend the
cooling time further \citep[see table 3 in ][]{chandra09}.

In the radiative regime, the existence of a cooled shell between the reverse
and forward shocks gives rise to additional X-ray absorption, and the column
density of the cool gas reduces with time as the SN expands
\citep{chevalier94,fransson96,chandra09}. With the high-quality spectra
collected by {\it XMM-Newton}, our result provides, for the first time, 
strong evidence for the decrease of absorption over the last decade (Table
\ref{combined_fits}), which agrees well with the model discussed above. We
further speculate that the non-detection of SN 1978K on 1980 January 2 could be
due to the large absorption by the cool shell.

\section*{Acknowledgements}
We thank the referee, Prof. Eric M. Schlegel, for the helpful
comments. We thank Drs Ming-Yu Ge and Qi-Rong Yuan for many valuable
suggestions. This work is supported by the National Natural Science Foundation
of China under grants 11303022, 11133002, 11573023, 11173016, 11673013 and
11433005, and by the Special Research Fund for the Doctoral Program of Higher
Education (grant No. 20133207110006).


\end{document}